\documentclass[aps,eqsecnum,nofootinbib,superscriptaddress,showpacs]{revtex4-2}
\def\be{\begin{eqnarray}}
\def\ee{\end{eqnarray}}
\usepackage[dvipdfmx]{graphicx}
\usepackage{amsfonts,bm}
\usepackage{amsmath,amsfonts,amssymb,bm}
\usepackage[dvips]{color}
\def\p{\partial}

\newcommand{\Exp}[1]{\left\langle~#1~\right\rangle}
\newcommand{\calT}{\mathcal{T}}
\newcommand{\bdyg}{\mathcal{G}}
\newcommand{\bdyR}{\mathcal{R}}
\newcommand{\bdyD}{\mathcal{D}}

\def\ben{\begin{equation}}
\def\een{\end{equation}}
\def\bena{\begin{eqnarray}}
\def\eena{\end{eqnarray}}

\begin{document}
\title{Semiclassical Rotating AdS Black Holes with Quantum Hair in Holography}

\author{Ryusei {\sc Hamaki}}\email[]{to69647@gmail.com}
\affiliation{%
{\it College of Systems Engineering and Science, 
Shibaura Institute of Technology, Saitama 330-8570, JAPAN
}}

\author{Kengo {\sc Maeda}}\email[]{maeda302@sic.shibaura-it.ac.jp}
\affiliation{%
{\it Faculty of Engineering,
Shibaura Institute of Technology, Saitama 330-8570, JAPAN}}

\begin{abstract}
In the context of the AdS/CFT duality, we study semiclassical stationary rotating AdS black holes with non-trivial quantum hair in three and five dimensions. We construct these solutions by perturbing the BTZ black hole and the five-dimensional Myers-Perry AdS black hole according to holographic semiclassical equations.
In the three-dimensional case, the vacuum expectation value of the stress-energy tensor diverges as 
$\sim 1/\lambda^n~(n=1,2)$ along a radial null geodesic as the affine parameter $\lambda$ approaches 
zero at the Cauchy horizon, depending on the type of perturbation. In the five-dimensional case, most hairy solutions exhibit strong divergences, either in the stress-energy tensor or in the parallelly propagated Riemann components, along the radial null geodesic crossing the Cauchy horizon. Nevertheless, there exists a specific class of semiclassical solutions that retain a 
$C^0$-regular Cauchy horizon, where perturbations remain bounded. 
For extremal black holes, the vacuum expectation value of the stress-energy tensor diverges along a radial null geodesic transverse to the event horizon in both three and five dimensions, even though all components of the perturbed metric vanish in this limit.

\end{abstract}
\maketitle

\section{Introduction}
Charged and rotating black holes have garnered significant attention in both the gravity and high-energy physics communities. A key topic of interest is the stability of the Cauchy horizon, which lies deep inside the event horizon. 
According to the strong cosmic censorship (SCC) conjecture, proposed by Penrose~\cite{Penrose1978}, the evolution of spacetime should be uniquely determined by regular initial data under physically reasonable conditions. In this framework, 
the presence of a Cauchy horizon is generally considered unphysical, as it would indicate a breakdown of determinism in general relativity. In classical general relativity, it is well established that the Cauchy horizon is generically unstable against perturbations~\cite{Simpson:1973ua, Brady:1995un}. However, the resulting singularity is weak in the sense that tidal distortions remain finite for an observer crossing the Cauchy horizon~\cite{Ori:1991zz}. Consequently, many 
$C^0$ extensions of the metric are possible beyond the Cauchy horizon, suggesting a potential violation of the SCC in classical theory.

Thus, it is natural to ask whether quantum effects can uphold the SCC conjecture. One tractable approach is to study quantum field theory in a fixed curved spacetime background. A quantized free scalar field has been extensively analyzed in various charged and rotating black hole backgrounds~\cite{Hiscock:1977qe, Hiscock:1980wr, Steif:1993zv, Dias:2019ery, Hollands:2019whz}.
Recent work, for instance, rigorously demonstrated that the vacuum expectation value of the stress-energy tensor diverges along a radial null geodesic as $\sim C\lambda^{-2}$, where 
$\lambda$ is the affine parameter, near the Cauchy horizon ($\lambda=0$) of a four-dimensional 
Reissner-Nordstr\"{o}m-de Sitter black hole~\cite{Hollands:2019whz}. In contrast, for the near-extremal 
rotating BTZ black hole, a three-dimensional vacuum solution with a negative cosmological constant~\cite{Banados:1992wn}, the vacuum expectation value remains finite, as shown in~\cite{Dias:2019ery}, due to the vanishing coefficient 
$C=0$~(see~\cite{Hollands:2019whz} for detailed analysis). These findings indicate that the coefficient 
$C$, which determines the formation of a strong curvature singularity at the Cauchy horizon, depends on the specific background spacetime. Therefore, it remains unclear whether quantum field theory in a fixed curved background can consistently support the SCC conjecture. An alternative scenario supporting the SCC conjecture could involve 
significant quantum modifications to the background geometry itself. As observed within a classical 
theory~\cite{Hartnoll:2020rwq}, one might expect that the backreaction destroys the Cauchy horizon of 
charged or rotating black holes. 

Another intriguing feature of charged or rotating black holes is the existence of extremal states with zero temperature. Motivated by the AdS/CFT duality~\cite{Maldacena:1997re}, recent studies have explored static or stationary perturbations of extremal Reissner-Nordstr\"{o}m-AdS black holes and rotating AdS black holes near their event horizons~\cite{Maeda:2011pk, Iizuka:2022igv, Horowitz:2022mly, Horowitz:2024kcx}.
As one approaches the horizon from the outside, the energy density or a curvature component in the parallelly propagated frame along a causal geodesic crossing the horizon diverges, despite the perturbations converging to zero and all scalar curvature invariants remaining finite. Such singularities are absent in non-extremal black holes with finite temperature. However, the curvature at the horizon grows indefinitely as the temperature decreases~\cite{Horowitz:2022mly}. This behavior suggests a potential violation of the weak cosmic censorship (WCC), which posits that a distant observer near future null infinity cannot witness the formation of a naked singularity.
Therefore, it becomes important to investigate whether quantum effects can prevent extremal black holes from developing curvature singularities at their event horizons.
 
In this paper, we investigate semiclassical rotating AdS black holes in both non-extremal and extremal cases. Given the absence of a complete theory of quantum gravity, the most tractable approach is the semiclassical procedure, where 
gravity is treated classically, while matter fields are treated quantum mechanically through the semiclassical Einstein equations. The source term in these equations is provided by the vacuum expectation value of the stress-energy tensor.
Previous studies have been limited to lower-dimensional semiclassical black hole solutions sourced by a quantized, free, massless scalar field~\cite{Casals:2016odj}, or to four-dimensional spherically symmetric solutions under the s-wave approximation of a massless, minimally coupled scalar field~\cite{Arrechea:2021ldl}. To extend these works, we explore semiclassical solutions for strongly coupled quantum fields with a gravity dual via the AdS/CFT duality.

Initially proposed in Ref.\cite{Compere:2008us}, the semiclassical equations on a $d$-dimensional boundary 
spacetime can be reformulated as mixed boundary conditions on a ($d+1$)-dimensional asymptotically AdS bulk spacetime by promoting the boundary metric to a dynamical field. Following this procedure, an analytical semiclassical (static) AdS black hole solution was obtained by introducing even-parity perturbations to the 
BTZ solution~\cite{Ishibashi:2023luz}. Extending this analysis to rotating cases, we construct 
three- and five-dimensional semiclassical rotating AdS black holes by perturbing the rotating BTZ solution 
and the five-dimensional Myers-Perry AdS solution~\cite{Hawking:1998kw}, respectively.

In the three-dimensional case, the non-extremal solution is obtained analytically by adding odd-parity perturbation to the static hairy solution~\cite{Ishibashi:2023luz} and applying a Lorentz boost. For the rotating solution, the vacuum stress-energy tensor diverges as $\sim \lambda^{-n}~(n=1,2)$ along a radial null geodesic with affine parameter 
$\lambda$ crossing the Cauchy horizon ($\lambda=0$), depending on the type of perturbation.
In the five-dimensional case, we derive three coupled second-order differential equations with six independent solutions. Interestingly, while the stress-energy tensor or the parallelly propagated Riemann components along the radial 
null geodesic strongly diverges as $\sim \lambda^{-2}$ near the Cauchy horizon for most parameter choices, 
it is possible to fine-tune the parameters to construct a specific hairy solution with a $C^0$-regular Cauchy horizon.

In the extremal case, the semiclassical extremal solution in three-dimensions is obtained analytically. 
The perturbation falls off as a power law, $\sim (r-r_0)^\gamma$, near the extremal horizon, $r=r_0$, with a very 
slow decay rate, i.~e.~, $0<\gamma<1$. This slow decay indicates the formation of a curvature singularity on the horizon, similar to the case of classical perturbations in extremal black holes~\cite{Maeda:2011pk, Iizuka:2022igv, Horowitz:2022mly, Horowitz:2024kcx}. This singular 
behavior also leads to the divergence of the vacuum expectation value of the stress-energy tensor 
along a radial null geodesic transverse to the horizon.
In the five-dimensional case, there are six independent mode solutions near the extremal horizon. 
We show that the power-law index $\gamma$ satisfies $\gamma<2$ for all parameters of the background 
Myers-Perry AdS solution, similar to previous semiclassical models~\cite{Arrechea:2021ldl}. 
This suggests that the formation of a curvature singularity on the extremal horizon is a generic feature of 
semiclassical extremal rotating black holes.

This paper is organized as follows: In section~\ref{sec:2}, we briefly review the holographic set up 
for constructing the semiclassical solutions. In section~\ref{sec:3} and \ref{sec:4}, we construct 
three and five-dimensional semiclassical rotating AdS black hole solutions. Section \ref{sec:5} is 
devoted to summary and discussions.


\section{Preliminaries: Holographic setting}
\label{sec:2}
We start with $d+1~(d=3,\,5)$-dimensional asymptotically AdS bulk spacetime with metric 
\begin{align}
\label{bulk_metric}
ds_{d+1}^2&=G_{MN}dX^MdX^N \nonumber \\
&=\Omega^{-2}(z)dz^2+g_{\mu\nu}(z,x)dx^\mu dx^\nu=\Omega^{-2}(z)(dz^2+ds_d^2), \nonumber \\  
ds_d^2&:=\tilde{g}_{\mu\nu}(z,x)dx^\mu dx^\nu, \qquad \Omega(z):=\frac{\ell}{L}\sin\frac{z}{\ell}, 
\end{align}
where $X^M=(z,\,x^\mu)$ and $L$~($\ell$) is the bulk~(boundary) AdS length and the AdS boundary is located 
at $z=0$ and $z=\pi$. The bulk action 
consists of the $d+1$-dimensional Einstein Hilbert action $S_\text{bulk}$, Gibbons-Hawking term $S_\text{GH}$, 
and the counter term $S_\text{ct}^{(d)}$ given by 
\begin{align}
\label{bulk_action}
& S_\text{bulk}=S_\text{EH}+S_\text{GH}+S_\text{ct}^{(d)} \nonumber \\
&=\int \frac{d^{d+1}X\sqrt{-G}}{16\pi G_{d+1}}\left(R(G)+\frac{d(d-1)}{L^2}  \right)
+\int \frac{d^dx\sqrt{-g}}{8\pi G_{d+1}}K+S_\text{ct}^{(d)}, 
\end{align}
where $G_d$ is the $d$-dimensional gravitational constant and $S^{(d)}_\text{ct}$ is given by
\begin{align}
\label{bulkct_3}
S^{(3)}_\text{ct}=-\int \frac{d^3x\sqrt{-g}}{16\pi G_4}\left(\frac{4}{L}+LR(g) \right), 
\end{align}
\begin{align}
\label{bulkct_5}
S^{(5)}_\text{ct}=-\int \frac{d^5x\sqrt{-g}}{16\pi G_6}\left(\frac{8}{L}+\frac{L}{3}R(g)
+\frac{L^3}{9}\left\{R_{\mu\nu}(g)R^{\mu\nu}(g)-\frac{5}{16}R^2(g)\right\}  \right). 
\end{align}

We consider $d$-dimensional spacetime with the conformal boundary metric 
$\bdyg_{\mu\nu}:=\lim_{z\to 0}\tilde{g}_{\mu\nu}(z,x)$ satisfying the semiclassical Einstein equations~\cite{Ishibashi:2023luz}.  According to the AdS/CFT dictionary~\cite{deHaro:2000vlm, Balasubramanian:1999re}, 
the vacuum expectation value of the boundary stress-energy tensor is given by the variation of the 
bulk action~(\ref{bulk_action}) with respect to the conformal boundary metric 
$\bdyg_{\mu\nu}$ as~\footnote{In detail derivation, see Ref.~\cite{Ishibashi:2023luz}.} 
\begin{align}
\label{vev_SE_tensor}
& \Exp{\calT_{\mu\nu}}
=\lim_{z\to 0}\frac{1}{8\pi G_{d+1}L}\Biggl{[\dfrac{L^2}{(d-2)\Omega^{d-2}}
\Bigl\{\tilde{K}\tilde{K}_{\mu\nu}-\dfrac{\tilde{g}_{\mu\nu}}{2}(\tilde{K}^{\alpha\beta}
\tilde{K}_{\alpha\beta}+\tilde{K}^2})\Bigr\}
\nonumber \\
&-L\tilde{g}_{\nu\rho}\left(\frac{L\Omega}{d-2}\frac{\p}{\p z}+1\right)
\frac{{\tilde{K}_\mu}^\rho-{\delta_\mu}^\rho\tilde{K}}{\Omega^{d-1}}
\nonumber \\
&+\frac{(d-1)\tilde{g}_{\mu\nu}}{(d-2)\Omega^d}\Bigl\{L^2\Omega^2\left(\frac{\Omega'}{\Omega}\right)'
+1-\frac{d-2}{2}(1-L\Omega')^2\Bigr\}\Biggr]+\tau^{(d)}_{\mu\nu}, 
\end{align}
where a prime denotes differentiation with respect to $z$ and the extrinsic curvature $\tilde{K}_{\mu\nu}$ of the 
conformal metric $\tilde{g}_{\mu\nu}$ is defined by  
\begin{align}
\label{def_tilextr}
\tilde{K}_{\mu\nu}=-\frac{1}{2}\p_z\tilde{g}_{\mu\nu}, 
\end{align}
and 
\begin{align}
\label{vev_SE_ct}
\tau^{(3)}_{\mu\nu}&=0, \nonumber \\
\tau^{(5)}_{\mu\nu}&=\frac{L^3}{72\pi G_6\Omega}\Biggl\{\left(\tilde{D}^2-\frac{5}{8}\tilde{R}\right)\tilde{R}_{\mu\nu}
-\frac{3}{8}\tilde{D}_\mu \tilde{D}_\nu \tilde{R}+2\tilde{R}_{\alpha\mu\beta\nu}\tilde{R}^{\alpha\beta}
\nonumber \\
&-\frac{\tilde{g}_{\mu\nu}}{2}
\left(\tilde{R}_{\alpha\beta}\tilde{R}^{\alpha\beta}+\frac{1}{4}\tilde{D}^2\tilde{R}
-\frac{5}{16}\tilde{R}^2\right)  \Biggr\}. 
\end{align}
Here, $\tilde{R}_{\mu\nu}$ is the Ricci tensor of the conformal metric $\tilde{g}_{\mu\nu}$. 

The bulk geometry~(\ref{bulk_metric}) is the solution of the $d+1$-dimensional bulk Einstein equations 
with a negative cosmological constant $\Lambda_{d+1}=-d(d-1)/(2L^2)$ if the $d$-dimensional 
metric $\tilde{g}_{\mu\nu}=\tilde{g}_{\mu\nu}(x)$ satisfies the vacuum Einstein equations with a 
negative cosmological constant $\Lambda_{d}=-(d-1)(d-2)/(2\ell^2)$. So, if the metric $\tilde{g}_{\mu\nu}$ 
corresponds to that of the BTZ black hole~\cite{Banados:1992wn} or the $5$-dimensional  
Myers-Perry AdS black hole~\cite{Hawking:1998kw}, the bulk Einstein equations are automatically satisfied. 
Furthermore, the vacuum expectation 
value of the stress-energy tensor~(\ref{vev_SE_tensor}) vanishes identically in both $d=3,\,5$ cases, due to 
the absence of the Weyl anomaly in odd dimensions. This implies that these bulk geometries are solutions to the semiclassical Einstein equations~\footnote{Here, for simplicity, we ignored higher curvature 
corrections in $d=5$ case.} on the boundary:
\begin{align}
\label{semi_eqs}
\bdyR_{\mu\nu}-\frac{\bdyR}{2}\bdyg_{\mu\nu}+\Lambda_d\,\bdyg_{\mu\nu}=8\pi G_d\Exp{\calT_{\mu\nu}}, 
\end{align}
where $\bdyR_{\mu\nu}$ is the Ricci tensor of the metric $\bdyg_{\mu\nu}$.  

We aim to find semiclassical solutions of Eqs.~(\ref{semi_eqs}) with non-zero expectation values, i.~e.~, 
$\Exp{\calT_{\mu\nu}}\neq 0$, by perturbing the background solution. To achieve this, 
we expand the metric $\tilde{g}_{\mu\nu}$ as 
\begin{align}
\label{exp_tildeg}
\tilde{g}_{\mu\nu}=\overline{\bdyg}_{\mu\nu}(x)+\epsilon h_{\mu\nu}(z,\,x)+\cdots, 
\end{align}
where $\overline{\bdyg}_{\mu\nu}(x)$ is the background metric of the BTZ black hole or the $5$-dimensional Myers-Perry AdS black hole, and $\epsilon$ is an infinitesimally small parameter.   
Under the metric ansatz $h_{\mu\nu}=\xi(z)H_{\mu\nu}(x)$, the perturbed bulk Einstein equations can 
be decomposed with a separation constant $m^2$ as
\begin{align}
\label{xi_equation}
\xi''+m^2\xi-\frac{(d-1)\Omega'}{\Omega}\xi'=0, 
\end{align}
\begin{align}
\label{H_equation}
\bar{\bdyD}^2H_{\mu\nu}+2\overline{\bdyR}_{\mu\alpha\nu\beta}H^{\alpha\beta}=m^2H_{\mu\nu},  
\end{align} 
where $\overline{\bdyD}_\mu$ and $\overline{\bdyR}_{\mu\nu\alpha\beta}$ are the covariant derivative 
and the Riemann tensor with respect to the background metric $\overline{\bdyg}_{\mu\nu}$, respectively, and 
we impose the transverse and traceless condition 
\begin{align}
\label{TT_cond}
{h_\mu}^\mu=h_{\mu\nu}\overline{\bdyg}^{\mu\nu}=0 \,, \quad \bar{\bdyD}^\nu h_{\mu\nu}=0 \,. 
\end{align}

The AdS boundary of the bulk spacetime~(\ref{bulk_metric}) consists of two portions: $z=0$, where the 
conformal boundary metric $\bdyg_{\mu\nu}$ satisfying the semiclassical Einstein equations, and $z=\pi$. 
According to the procedure~\cite{Ishibashi:2023luz}, we shall impose the Dirichlet boundary condition 
at $z=\pi$. As shown in Ref.~\cite{Ishibashi:2024fnm}, the solution of Eq.~(\ref{xi_equation}) 
is expressed by the hypergeometric function as
\begin{align}
\label{xi_sol}
\xi(y)=C(1-y)^\frac{d}{2}F\left(\frac{1}{2}-p,\,\frac{1}{2}+p,\,1+\frac{d}{2};1-y  \right), 
\qquad y:=\frac{1-\cos\frac{z}{\ell}}{2}, 
\end{align}
where $C$ is a constant and $p$ is defined by
\begin{align}
\label{def_p}
p=\sqrt{\frac{(d-1)^2}{4}+\hat{m}^2} 
\end{align}
with $\hat{m}^2:=m^2\ell^2$. 
As shown in the perturbations of the maximally symmetric AdS 
spacetimes~\cite{Ishibashi:2023luz, Ishibashi:2024fnm}, 
the perturbed semiclassical Eqs.~(\ref{semi_eqs}) reduce to an algebraic equation characterized by 
the dimensionless constant $\gamma_d$, defined by 
\begin{align} 
\gamma_d := \frac{G_dL}{\pi G_{d+1}}\left(\frac{L}{\ell} \right)^{d-2}. 
\end{align}
For $d=3$, the first order perturbation of Eqs.~(\ref{semi_eqs}) using Eqs.~(\ref{vev_SE_tensor}), (\ref{vev_SE_ct}), and 
(\ref{xi_sol}) results in the algebraic equation~\cite{Ishibashi:2023luz}
\begin{align}
\label{algebra_3D}
\gamma_3=\frac{\tan \pi p}{\pi p},  
\end{align}
which allows a hairly solution in the range
\begin{align}
\label{mass_inequality_3D}
\gamma_3>1, \qquad -1<\hat{m}^2<-\frac{3}{4}.  
\end{align}

For $d=5$, the perturbed semiclassical Eqs.~(\ref{semi_eqs}) in the Myers-Perry AdS black hole 
background reduce to the algebraic equation~(\ref{algebra_5D}), which is identical to that derived 
in the perturbation of the $5$-dimensional maximally symmetric AdS solution~\cite{Ishibashi:2024fnm}. 
The solution was found for a negative mass satisfying the inequality
\begin{align}
\label{mass_inequality_5D}
-\frac{15}{4}<\hat{m}^2<-\frac{7}{4}  
\end{align}  
when $\gamma_5$ is larger than a critical value~\cite{Ishibashi:2024fnm}. 
In the following sections, we shall focus on the solutions of (\ref{H_equation}) under the inequalities,  (\ref{mass_inequality_3D}) and (\ref{mass_inequality_5D}), ensuring that the algebraic 
equations~(\ref{algebra_3D}) and (\ref{algebra_5D}) are also satisfied. 

\section{$3$-dimensional rotating AdS black holes}
\label{sec:3}
In this section, we investigate $3$-dimensional rotating AdS black holes by solving Eqs.~(\ref{H_equation}) 
for a negative mass satisfying the inequality (\ref{mass_inequality_3D}) in the background of the rotating BTZ 
solution~\cite{Banados:1992wn}. In the non-extremal case, an analytic solution is constructed by adding an odd-parity 
perturbation to the static solution~\cite{Ishibashi:2023luz} and by applying a Lorentz boost to the perturbed solution. 
In the extremal case, we also obtain an analytic semiclassical solution in which $H_{\mu\nu}$ in Eqs.~(\ref{H_equation}) 
converges to zero near the horizon. Nevertheless, we demonstrate that the expectation value of the stress-energy 
tensor diverges along a null geodesic crossing the event horizon. 

\subsection{The non-extremal solution}
We start with the metric ansatz for the non-extremal AdS black hole solution,  
\begin{align} 
\label{rot_BTZ_ansatz}
& ds_3^2=\bdyg_{\mu\nu}dx^\mu dx^\nu
=(\overline{\bdyg}_{\mu\nu}+\epsilon H_{\mu\nu})dx^\mu dx^\nu \nonumber \\
&=-\frac{r_h^2g(u)}{\ell^2(1+\sinh^2\beta\, u)u}(1+\epsilon K(u))dt^2+\frac{\ell^2}{4u^2g(u)}(1+\epsilon U(u))du^2
\nonumber \\
&+\frac{r_h^2(1+\sinh^2\beta\, u)}{u}(1+\epsilon Z(u))
\left[\left(-\frac{\cosh\beta\sinh\beta\, u}{\ell(1+\sinh^2\beta\, u)}+\epsilon\Omega(u)\right)dt+d\varphi \right]^2, 
\nonumber \\
& g(u)=1-u, 
\end{align}
where $0\le u\le \infty$, $0\le \varphi\le 2\pi$, $\beta~(0<\beta<\infty)$ is a boost parameter, and 
the event horizon~(the AdS boundary) is located at $u=1$~($u=0$). 
It can be verified that the metric with $\epsilon=0$ reduces to standard form of the rotating BTZ 
solution~\cite{Banados:1992wn} through the coordinate transformation 
\begin{align}
u=\frac{r_h^2}{r^2-r_h^2\sinh^2\beta}, 
\end{align}
where the radii of the event horizon and the Cauchy horizon correspond to $r=r_h\cosh\beta$, and 
$r=r_h\sinh\beta$~($u=\infty$), respectively.   

Under the Lorentz boost, 
\begin{align}
\label{boost}
\begin{pmatrix}
   t/\ell  \\
   \varphi 
\end{pmatrix}
=\begin{pmatrix}
   \cosh\beta & \sinh\beta \\
    \sinh\beta & \cosh\beta 
\end{pmatrix}
\begin{pmatrix}
   \hat{t}/\ell  \\
   \hat{\varphi} 
\end{pmatrix}, 
\end{align}
the metric~(\ref{rot_BTZ_ansatz}) reduces to 
\begin{align} 
\label{BTZ_static_u}
ds_3^2=-\frac{r_h^2g}{\ell^2 u}(1+\epsilon T(u))d\hat{t}^2
+\frac{\ell^2}{4u^2g}(1+\epsilon U(u))du^2+\frac{r_h^2}{u}(1+\epsilon R(u))d\hat{\varphi}^2
-\frac{2\epsilon r_h^2}{u}S(u)d\hat{t}d\hat{\varphi}, 
\end{align}
where 
\begin{align} 
\label{KLOmega}
& K=\frac{gR+T+(T-gR)\cosh(2\beta)-2\ell\sinh(2\beta) S}{2+u(\cosh(2\beta)-1)}, \nonumber \\
& Z=\frac{R+gT+(R-gT)\cosh(2\beta)+2\ell\sinh(2\beta) S}{2+u(\cosh(2\beta)-1)}, \nonumber \\
& \Omega=-2\frac{\ell S\{u+(2-u)\cosh(2\beta)\}+g(R-T)\sinh(2\beta)}{\ell\{2+u(\cosh(2\beta)-1)\}^2}. 
\end{align}
In this metric form~(\ref{BTZ_static_u}), the perturbation for $S$ is decoupled from the other variables, $(U,\,T,\,R)$, 
since they correspond to odd and even-parity mode perturbations under 
$\hat{\varphi}\leftrightarrow -\hat{\varphi}$ transformation. 

For the even-parity mode, one obtains a master equation of $U$~\cite{Ishibashi:2023psu}, 
and $R$, $T$ are determined by Eq.~(\ref{TT_cond}) as 
\begin{align} 
\label{3dim_traceless}
R=-(T+U), 
\end{align}
\begin{align} 
\label{3dim_transverse}
T=-\left(2-\frac{3}{u}\right)U-2(1-u)U', 
\end{align}
\begin{align} 
\label{master_U}
gU''-\frac{2}{u}U'+\frac{8-\hat{m}^2}{4u^2}U=0, 
\end{align}
where a prime denotes differentiation with respect to $u$. 
For the odd-parity mode, the equation of $S$ becomes 
\be 
\label{S}
gS''-\frac{\hat{m}^2}{4u^2}S=0. 
\ee
By imposing regularity on the event horizon, $u=1$, the solutions are analytically given by 
\begin{align}
\label{sol_3dim_US}
& U=c_1\zeta(u), \qquad 
S=\frac{c_2g}{\ell u}\zeta(u), \nonumber \\
& \zeta(u):=u^\frac{3-p}{2}F\left(\frac{1-p}{2}, \frac{3-p}{2}, 2; 1-u  \right),
\end{align}
where $c_1$, $c_2$ are arbitrary constants, $F$ is the hypergeometric function, and 
$p$ is the parameter defined by 
$p=\sqrt{1+\hat{m}^2}$~(\ref{def_p}). So, by Eqs.~(\ref{KLOmega}), (\ref{3dim_traceless}), and 
(\ref{3dim_transverse}), the general rotating AdS black hole solution with quantum 
hair is described by the two parameters, $c_1$ and $c_2$, as follows:  
\begin{align} 
\label{sol_KLOmega}
& K=\frac{c_1\{u(2-u)+(u^2-6u+6)\cosh(2\beta)\}-2c_2(1-u)\sinh(2\beta)}{u(2-u+u\cosh(2\beta))}\zeta  
-\frac{2c_1g\{u+(2-u)\cosh(2\beta)\}}{2-u+u\cosh(2\beta)}\zeta', \nonumber \\
& L=-2\frac{c_1\{u(2-u)+(u^2-3u+3)\cosh(2\beta)\}-c_2(1-u)\sinh(2\beta)}{u(2-u+u\cosh(2\beta))}\zeta  
+\frac{2c_1g\{u+(2-u)\cosh(2\beta)\}}{2-u+u\cosh(2\beta)}\zeta', \nonumber \\
& \Omega=-2g\frac{c_2\{u+(2-u)\cosh(2\beta)\}-3c_1(2-u)\sinh(2\beta)}{\ell u(2-u+u\cosh(2\beta))^2}\zeta  
-\frac{8c_1g^2\sinh(2\beta)}{\ell(2-u+u\cosh(2\beta))^2}\zeta'. 
\end{align}
Expanding $\zeta$~(\ref{sol_3dim_US}) near the null infinity, $u=0$, one obtains 
\begin{align}
U=O(u^\frac{3-p}{2}), \qquad R,\,T,\,S=O(u^\frac{1-p}{2}). 
\end{align}
This guaranties the asymptotic boundary condition that $K$, $Z$, and $\Omega$ decay to zero at $u=0$ for 
$0<p<1/2$~(\ref{mass_inequality_3D}). 

Now, consider the behavior of the vacuum expectation value of the perturbed stress-energy tensor $\delta\Exp{T_{\mu\nu}}$ 
near the Cauchy horizon, $u=\infty$. Near the AdS boundary $z=0$, the metric $h_{\mu\nu}$ in (\ref{exp_tildeg}) can be 
expanded as 
\begin{align}
h_{\mu\nu}(z,\,x)=h^{(0)}_{\mu\nu}(x)+\left(\frac{z}{L}\right)^2h^{(2)}_{\mu\nu}(x)+
\left(\frac{z}{L}\right)^3h^{(3)}_{\mu\nu}(x)+\cdots, 
\end{align}
and the coefficient $h^{(3)}_{\mu\nu}(x)$ corresponds to $\delta\Exp{T_{\mu\nu}}$~\cite{Ishibashi:2023luz}. 
Consequently, under the metric ansatz, $h_{\mu\nu}=\xi(z)H_{\mu\nu}(x)$, we obtain the perturbed 
stress-energy tensor 
in the coordinate system~(\ref{BTZ_static_u}) as 
\begin{align}
\label{stress_energy_BTZ}
& \delta\Exp{T_{\mu\nu}}dx^\mu \otimes dx^\nu\sim\frac{3\epsilon}{16\pi G_4L}\delta(ds_3^2) 
\nonumber \\
&=\frac{3\epsilon}{16\pi G_4L}\left(-\frac{r_h^2g}{\ell^2 u}T(u)d\hat{t}^2
+\frac{\ell^2}{4u^2g}U(u)du^2+\frac{r_h^2}{u}R(u)d\hat{\varphi}^2
-\frac{2r_h^2}{u}S(u)d\hat{t}d\hat{\varphi}\right). 
\end{align}
  
Let us denote the tangent vector along an affine parameterized null geodesic crossing the Cauchy horizon 
as $V^\mu$. Decomposing the tangent null vector $V^\mu$ into the background 
$\overline{V}^\mu$ and the first order deviation $\delta V^\mu$,~i.~e.~, 
$V^\mu=\overline{V}^\mu+\delta V^\mu$, one obtains 
\begin{align} 
\label{pert_stress_energy}
\delta (\Exp{T_{\mu\nu}}V^\mu V^\nu)
=2\overline{\Exp{T_{\mu\nu}}}\delta V^\mu \overline{V}^\nu
+\delta\Exp{T_{\mu\nu}}\overline{V}^\mu\overline{V}^\nu
=\delta\Exp{T_{\mu\nu}}\overline{V}^\mu\overline{V}^\nu, 
\end{align}
where we used the fact the background stress-energy tensor $\overline{\Exp{T_{\mu\nu}}}$ is zero in the second equality. 

The background null vector $\overline{V}^\mu$ in the metric~(\ref{BTZ_static_u}) is expressed by two parameters 
${\cal E}$ and ${\cal P}$,  
\begin{align}
\overline{V}^\mu=u\left(\frac{2r_h}{\ell}\sqrt{({\cal E}^2-g{\cal P}^2)u}\frac{\partial}{\partial u}
+\frac{\ell {\cal E}}{g}\frac{\partial}{\partial \hat{t}}
+{\cal P}\frac{\partial}{\partial \hat{\varphi}}\right)^\mu. 
\end{align}
Contracting $\delta\Exp{T_{\mu\nu}}$ in Eq.~(\ref{stress_energy_BTZ}) with $\overline{V}^\mu$, 
one obtains 
\begin{align}
\label{null_energy_BTZ}
\delta (\Exp{T_{\mu\nu}}V^\mu V^\nu)\sim \frac{3\epsilon}{16\pi G_4L}\frac{r_h^2u}{g}
\left[{\cal P}^2g(R-U)-2\ell {\cal E}{\cal P} S+{\cal E}^2(U-T)   \right]. 
\end{align}

Expanding $\zeta$ in Eq.~(\ref{sol_3dim_US}) as a series in $1/u$, 
the asymptotic behavior of $\zeta$ near the CH~($u=\infty$) is given by 
\begin{align}
\zeta\simeq \alpha u, \qquad 
\alpha:=\frac{4\cos\left(\frac{\pi p}{2} \right)}{\pi(1-p^2)}. 
\end{align}
Inserting this into Eqs.~(\ref{3dim_traceless}), (\ref{3dim_transverse}), and (\ref{sol_3dim_US}), 
one obtains the leading behaviors of $U$, $T$, $R$, and $S$ near the CH as
\begin{align}
U\simeq -R\simeq c_1\alpha u, \quad T=O(1), \quad S\simeq -\frac{c_2\alpha}{\ell} u,   
\end{align}
and, by Eqs.~(\ref{null_energy_BTZ}), the expectation value of the stress-energy tensor diverges as 
\begin{align}
\delta (\Exp{T_{\mu\nu}}V^\mu V^\nu)\sim -\frac{3\epsilon\alpha r_h^2}{8\pi G_4L}(c_1{\cal P}^2u^2-c_2{\cal E}{\cal P} u)
\simeq c_1(\lambda^{-2})+c_2(\lambda^{-1}). 
\end{align}
Here, $\lambda$ is the affine parameter along the null geodesic, with $\lambda=0$ corresponding to the location of 
the CH, $u=\infty$. 
For the even-parity solution with $c_1\neq 0, \,c_2=0$, the expectation value of the stress-energy tensor 
strongly diverges, supporting the SCC conjecture. In contrast, for the odd-parity solution with $c_1=0, \,c_2\neq 0$, the 
divergence is weaker in the sense that the distance $\eta$ between the two neighboring null geodesics remains finite at 
the Cauchy horizon. This is because integrating $\delta (\Exp{T_{\mu\nu}}V^\mu V^\nu)$ twice with respect to 
$\lambda$ results in a finite value at the CH~\cite{Ori:1991zz}. Thus, a $C^0$ extension of the metric beyond 
the CH is possible.  

\subsection{The extremal solution}
\label{extremal_BTZ}
The extremal rotating black hole corresponds to the limit $\beta\to \infty$ in the Lorentz 
boost~(\ref{boost}). So, instead of the metric ansatz~(\ref{rot_BTZ_ansatz}),  
we consider the following metric ansatz for the extremal rotating solution:
\begin{align}
\label{ext_rotating_BTZ} 
& ds^2=-\frac{r_0^2}{\ell^2u}(1-u)^2(1+\epsilon T(u))dt^2
+\frac{\ell^2(1+\epsilon U(u))}{4u^2(1-u)^2}du^2 \nonumber \\
&+\frac{r_0^2}{u}(1+\epsilon R(u))\left\{\left(-\frac{u}{\ell}+\epsilon S(u)\right)dt+d\varphi \right\}^2, 
\end{align}
where $\epsilon=0$ corresponds to the extremal BTZ black hole~\footnote{The coodinate transformation 
$u=r_0^2/r^2$ produces the standard form of the extremal BTZ black hole metric with horizon radius 
$r_0$~\cite{Banados:1992wn}.}. 
Eqs.~(\ref{H_equation}) and (\ref{TT_cond}) lead to 
\begin{align} 
\label{ext_BTZ_R}
R=-(T+U), 
\end{align}
\begin{align}
\label{ext_BTZ_S}
S=\frac{1-u}{2\ell u}\left\{2(1-u)uU'-(3-u)U+2uT  \right\}, 
\end{align}
\begin{align} 
\label{ext_BTZ_U}
(1-u)^2U''-\frac{2}{u}(1-u^2)U'+\frac{8-\hat{m}^2}{4u^2}U=0, 
\end{align}
\begin{align} 
\label{ext_BTZ_T}
& (1-u)^2T''-\frac{\hat{m}^2}{4u^2}T+\frac{2(1-u^2)}{u}U'+\frac{\hat{m}^2-4}{2u^2}\,U=0.  
\end{align}

The general solution of Eq.~(\ref{ext_BTZ_U}) is 
expressed by the parameter $p=\sqrt{1+\hat{m}^2}$~(\ref{def_p}) as
\begin{align} 
\label{sol:ext_BTZ_U}
U=a_1\left(\frac{1-u}{u} \right)^{-\frac{3+p}{2}}+a_2\left(\frac{1-u}{u} \right)^{-\frac{3-p}{2}}. 
\end{align}
Each mode function in~(\ref{sol:ext_BTZ_U}) diverges at the extremal horizon, $u=1$ 
under the inequality~(\ref{mass_inequality_3D}), $0<p<1/2$, and therefore, $a_1=a_2=0$, i.~e., $U=0$. 
Substituting $U=0$ into Eqs.~(\ref{ext_BTZ_R}), (\ref{ext_BTZ_S}), 
and (\ref{ext_BTZ_T}), one obtains 
\begin{align} 
\label{sol:ext_BTZ}
& T=b_1u^\frac{1-p}{2}(1-u)^\frac{1+p}{2}+b_2u^\frac{1+p}{2}(1-u)^\frac{1-p}{2}, \nonumber \\
& R=-T, \qquad S=\frac{(1-u)T}{\ell}. 
\end{align}
In the range $0<p<1/2$, the perturbation goes to zero towards the AdS boundary, $u=0$, which is the same as the 
non-extremal case~(\ref{sol_3dim_US}).  It also converges to zero at the event horizon, $u=1$. Nevertheless, 
we can show that the stress-energy tensor along null geodesics crossing the event 
horizon~(\ref{pert_stress_energy}) diverges. Let us consider a null geodesic on the background BTZ black hole
with tangent vector $\overline{V}^\mu$,  
\begin{align} 
\label{ex_BTZ_null_tangent}
\overline{V}^\mu=
{\cal E}u\left(\frac{\ell^2}{2r_0(1-u)^2}\frac{\p}{\p t}+\sqrt{u}\frac{\p}{\p u}
+\frac{\ell u}{2r_0(1-u)^2}\frac{\p}{\p \varphi}\right)^\mu, 
\end{align}
where ${\cal E}$ is a positive constant. 
By Eqs.~(\ref{sol:ext_BTZ}) and (\ref{ex_BTZ_null_tangent}), we obtain 
\begin{align} 
& \delta(\Exp{T_{\mu\nu}}V^\mu V^\nu)=\delta\Exp{T_{\mu\nu}}\overline{V}^\mu\overline{V}^\nu
\sim \frac{3\epsilon}{16\pi G_4L} H_{\mu\nu}\overline{V}^\mu\overline{V}^\nu \nonumber \\
&\sim-\epsilon\frac{3{\cal E}^2\ell^2 uT(u)}{64\pi G_4L(1-u)^2}=\epsilon\, O\left( (1-u)^{-\frac{3+p}{2}}\right)
=\epsilon\, O\left( (-\lambda)^{-\frac{3+p}{2}}\right), 
\end{align}
where $\lambda$ is the affine parameter of the null geodesic with $\lambda\sim -(1-u)$.  
This singular behavior is similar to the semiclassical result in the extremal Reissner-Nordstr\"{o}m 
black hole~\cite{Arrechea:2021ldl}. 

\section{$5$-dimensional rotating AdS black holes}
\label{sec:4}
In this section we construct $5$-dimensional semiclassical rotating AdS black hole solutions by solving 
Eqs.~(\ref{H_equation}) and (\ref{TT_cond}). In general, it is difficult to solve these equations in the background 
of rotating AdS black holes because the perturbations are partial differential equations. One notable exception is 
the odd-dimensional Myers-Perry~(MP) AdS black hole~\cite{Hawking:1998kw} with 
equal angular momenta, which is known to be cohomogeneity-one solution of the vacuum 
Einstein equations with a negative cosmological constant. In this case, the equations become coupled 
ordinary differential equations, making it possible to analyze the singular behavior of the Cauchy horizon 
both analytically and numerically. Hereafter, we construct the solutions of Eqs.~(\ref{H_equation}) 
and (\ref{TT_cond}) for the $5$-dimensional MP AdS black hole with equal angular momenta in 
both non-extremal and extremal cases.   

\subsection{The non-extremal solution}
We consider the following metric ansatz for perturbation of the $5$-dimensional MP AdS black hole 
with equal angular momenta as 
\begin{align}
\label{metric_ansatz_5MP}
& ds^2=-\frac{\hat{r}_0^2\eta}{uh}(1+\epsilon T(u))dt^2+\frac{\ell^2}{4u^2\eta}(1+\epsilon U(u))du^2
+\frac{\hat{r}_0^2\ell^2}{4u}(1+\epsilon R(u))(d\theta^2+\sin^2\theta d\varphi^2)
\nonumber \\
&+\frac{\hat{r}_0^2\ell^2h}{u}(1+\epsilon Z(u))
\left(d\psi+\frac{\cos\theta}{2}d\varphi-(\Omega+\epsilon S(u))dt\right)^2, \nonumber \\ 
& \eta=(1-u)(1-\kappa^2 u)\left(\frac{u}{\hat{r}_0^2}+u+\kappa^2 u+1\right), \nonumber \\
& h=1+\kappa^2\{1+(1+\kappa^2)\hat{r}_0^2\}u^2, \qquad 
\Omega=\frac{M^2a}{\hat{r}_0^4h\ell^4}u^2, \nonumber \\
& M=\ell\hat{r}_0\sqrt{(1+\kappa^2)(1+\hat{r}_0^2)(1+\kappa^2\hat{r}_0^2)}, \nonumber \\
& a=\ell \kappa\hat{r}_0
\sqrt{\frac{1+(1+\kappa^2)\hat{r}_0^2}{(1+\kappa^2)(1+\hat{r}_0^2)(1+\kappa^2\hat{r}_0^2)}}, 
\end{align}
where each surface on $t=$const. and $u=$const. is a homogenously squashed $S^3$, which consists of 
$S^1$ fibers over $S^2$~(the agular coordinate ranges are $0\le \psi\le 2\pi$, $0\le \theta\le \pi$, 
and $0\le \varphi\le 2\pi$), and $\hat{r}_0~(=r_0/\ell)$ is the horizon radius on $S^2$ 
normalized by $\ell$. Here, the radial coordinate $u$ is defined such that the event horizon and the 
Cauchy horizon are located at $u=1$ and $u=1/\kappa^2$ in the range $0<\kappa\le 1$. 
Note that the $\epsilon=0$ background metric reduces to the standard form of the $5$-dimensional 
MP AdS black hole~\cite{Hawking:1998kw} with equal angular momenta under the coordinate transformation $u=r_0^2/r^2$.   

Under the metric ansatz, Eqs.~ (\ref{H_equation}) and (\ref{TT_cond}) are decomposed into the three coupled 
second order ordinary differential equations for $U$, $T$, and $Z$ 
\begin{align} 
\label{eq_MP_U}
& U''+\left(\frac{2\eta'}{\eta}-\frac{5}{u}\right)U'
+\left(\frac{h'^2}{h^2}+\frac{h}{\hat{r}_0^2u\eta}-\frac{h'\eta'}{h\eta}-\frac{\ell^2h^2\Omega'^2}{\eta} \right)T
+\left[-\frac{h}{\hat{r}_0^2u\eta}+\frac{1}{u^2\eta}(2\eta-2-u\eta') \right]Z
\nonumber \\
&+\left[\frac{8}{u^2}-\frac{h'^2}{h^2}-\frac{\hat{m}^2}{4u^2\eta}+\left(\frac{h'}{h}-\frac{4}{u}\right)\frac{\eta'}{\eta}
+\frac{\ell^2h^2\Omega'^2}{\eta}  \right]U=0,   
\end{align}
\begin{align} 
\label{eq_MP_T}
& T''+\left(\frac{h'}{h}-\frac{1}{u}\right)T'+\left(\frac{4}{u}+\frac{h'}{h}-\frac{2\eta'}{\eta}  \right)U'-\frac{h'}{h}Z'
+\left[\frac{h}{\hat{r}_0^2u\eta}+\frac{1}{u^2\eta}(2-2\eta+u\eta') \right]Z \nonumber \\
& +\left[-\frac{h'^2}{h^2}+\frac{1}{\eta}\left(-\frac{\hat{m}^2}{4u^2}-\frac{h}{\hat{r}_0^2u}
+\frac{h'\eta'}{h}+\ell^2h^2\Omega'^2 \right)    \right]T \nonumber \\
&+\left[-\frac{10}{u^2}+\frac{2h'^2}{h^2}
+\frac{1}{\eta}\left(\frac{h}{\hat{r}_0^2u}-\frac{2h'\eta'}{h}+\frac{4+\hat{m}^2+10u\eta'}{2u^2}
-2\ell^2h^2\Omega'^2    \right)   \right]U=0, 
\end{align}
\begin{align} 
\label{eq_MP_Z}
& Z''+\left(\frac{\eta'}{\eta}+\frac{h'}{h}-\frac{1}{u}  \right)Z'+\left(\frac{\eta'}{\eta}-\frac{h'}{h}-\frac{2}{u}  \right)U'
+\left(\frac{\eta'}{\eta}-\frac{h'}{h}  \right)T'
+\left(\frac{h'^2}{h^2}-\frac{h'\eta'}{h\eta}-\frac{\ell^2h^2\Omega'^2}{\eta}    \right)T \nonumber \\
&+\left[\frac{8}{u^2}-\frac{h'^2}{h^2}+\frac{1}{\eta}\left(-\frac{2h}{\hat{r}_0^2u}+\frac{h'\eta'}{h}
-\frac{8u\eta'+8+\hat{m}^2}{2u^2}+\ell^2h^2\Omega'^2 \right)   \right]U \nonumber \\
&-\left[\frac{3h}{\hat{r}_0^2u\eta}+\frac{1}{u^2\eta}\left(2+\frac{\hat{m}^2}{4}-2\eta+u\eta' \right) \right]Z=0, 
\end{align}
and $S$ and $R$ are determined by 
\begin{align} 
\label{eq_MP_S}
S=\frac{5\eta-u\eta'}{2\ell^2uh^2\Omega'}U+\frac{h\eta'-\eta h'}{2\ell^2h^3\Omega'}T
+\frac{\eta h'}{2\ell^2h^3\Omega'}Z-\frac{\eta U'}{\ell^2h^2\Omega'}, 
\end{align}
\begin{align} 
\label{eq_MP_R}
R=-\frac{U+T+Z}{2}, 
\end{align}
where a prime denotes differentiation with respect to $u$. 
Let us consider the asymptotic solution near the AdS boundary, $u=0$. Introducing the new variables, 
$P=uU'$, $Q=uZ'$, $R=uT'$, and $\bm{V}=(U, Z, T, P, Q, R)$, Eqs.~(\ref{eq_MP_U}), (\ref{eq_MP_T}), and 
(\ref{eq_MP_Z}) are transformed into the six coupled first order differential equations near $u=0$, 
\begin{align}
\label{matrix_asymp}
u\frac{d{\bm V}}{du}\simeq A{\bm V}, \qquad 
A:=
\begin{pmatrix}
0 & 0 & 0  & 1  & 0  &  0 \\
0 & 0 & 0 & 0 & 1 & 0 \\
0 & 0 & 0 & 0 & 0 & 1 \\
-\left(8-\frac{\hat{m}^2}{4}\right) & 0 & 0 & 6 & 0 & 0 
\\
\frac{\hat{m}^2}{2}-4 & \frac{\hat{m}^2}{4} & 0 & 2 & 2 & 0 \\
 8-\frac{\hat{m}^2}{2} & 0 & \frac{\hat{m}^2}{4} & -4 & 0 & 2 
\end{pmatrix}. 
\end{align}
Substituting ${\bm V}=u^\zeta {\bm K}$ into (\ref{matrix_asymp}), we arrive at the eigenvalue 
equation $|A-\zeta I|=0$, where $I$ is the identity matrix and $\zeta$ represents the eigenvalue. The solutions are given by 
\begin{align}
\label{eigensystem_asymp}
& \zeta=3\pm\frac{p}{2}, \qquad 
{\bm K}_{1\pm}
=\left(-\frac{8}{p^2\pm 8p+12}, \,\mp\frac{2p}{p^2\pm 8p+12},\, \pm\frac{2}{p\pm 6}, \, \mp\frac{4}{p\pm 2}, \,
-\frac{p}{p\pm 2},\,1   \right), \nonumber \\
& \zeta=1\pm \frac{p}{2}, \qquad {\bm K}_{2\pm}=\left(0,\,0,\,1,\,0,\,0,\,1\pm\frac{p}{2} \right), \qquad  
{\bm K}_{3\pm}=\left(0,\,1,\,0,\,0,\,1\pm\frac{p}{2},\,0\right). 
\end{align}
By the inequality~(\ref{mass_inequality_5D}), the range of $p$ is $1/2<p<3/2$. Within this range, all the variables, 
$U$, $Z$, $T$, and $R$ converge to zero for all eigenvalues $\zeta$ considered above. Additionally, we can show 
that $S$ in Eq.~(\ref{eq_MP_S}) also converges to zero as follows. 

Among all modes, ${\bm K}_{2-}$ and ${\bm K}_{3-}$ exhibit the slowest decay at infinity. 
For the ${\bm K}_{2-}$ mode, using Eqs.~(\ref{eq_MP_U}), (\ref{eq_MP_T}), and (\ref{eq_MP_Z}), 
$U$, $Z$, $T$ can be expanded as a series in $u$ as 
\begin{align}
\label{asymp_t_mode}
U\simeq-\frac{u^{2-\frac{p}{2}}}{(1+p)\hat{r}_0^2}, \qquad T\simeq u^{1-\frac{p}{2}}, \qquad 
Z=O\left(u^{2-\frac{p}{2}}\right). 
\end{align}  
Substituting Eq.~(\ref{asymp_t_mode}) into Eq.~(\ref{eq_MP_S}) and using $\Omega=O(u^2)$ and $h'=O(u)$, we 
find that $S=O(u^{1-p/2})$, which converges to zero in the range $1/2<p<3/2$. 
Similarly, for the ${\bm K}_{3-}$ mode, using Eqs.~(\ref{eq_MP_U}), (\ref{eq_MP_T}), and (\ref{eq_MP_Z}), one finds 
\begin{align}
\label{asymp_z_mode}
Z=O\left(u^{1-\frac{p}{2}}\right), \qquad U=T=O\left(u^{3-\frac{p}{2}}\right), 
\end{align}
leading to the same asymptotic behavior, $S=O(u^{1-p/2})$. 

One can also derive the asymptotic solutions of Eqs.~(\ref{eq_MP_U}), (\ref{eq_MP_T}), and (\ref{eq_MP_Z}) near the 
event horizon~(Cauchy horizon) by transforming them into the coupled first order differential equations near $u_\ast=1~(u_\ast=1/\kappa^2)$. 
Introducing variables $P$, $Q$, and $R$ as $P=(u_\ast-u)U'$, $Q=(u_\ast-u)Z'$, $R=(u_\ast-u)T'$, and  
setting ${\bm V}$ as $\bm{V}=(U, Z, T, P, Q, R):=(u_\ast-u)^\zeta {\bm K}$, the equations become
\begin{align}
(u_\ast-u)\frac{d{\bm V}}{du}\simeq A{\bm V}, \qquad 
\label{non-extremal_NH_A}
A:=
\begin{pmatrix}
0 & 0 & 0  & 1  & 0  &  0 \\
0 & 0 & 0 & 0 & 1 & 0 \\
0 & 0 & 0 & 0 & 0 & 1 \\
0 & 0 & 0 & 1 & 0 & 0 
\\
0 & 0 & 0 & 1 & 0 & 1 \\
0 & 0 & 0 & -2 & 0 & -1
\end{pmatrix}  
.
\end{align}
The solutions of the eigenvalue equation $|A+\zeta I|=0$ are $\zeta=\pm 1$ and $0$, 
where the eigenvectors of $\zeta=\pm 1$ are given by 
\begin{align}
\label{non_trivial_eigen}
\zeta=-1: \,\,\, {\bm K}_-=(1,\,0,\,-1,\,1,\,0,\,-1), \qquad  
\zeta=1: \,\,\, {\bm K}_+=(0, 1, -1, 0, -1, 1). 
\end{align}
The zero mode solutions consist of two independent regular solutions and one logarithmic divergent mode solution, 
$Z\sim \ln(u_\ast-u)$, as derived from Eq.~(\ref{eq_MP_Z}). 

By imposing regularity at the event horizon, and expanding $U$, $T$, and $Z$ as 
\begin{align}
\label{expantion_EH}
& U=u_0-u_1(1-u)+u_2(1-u)^2+\cdots, \nonumber \\ 
& T=t_0-t_1(1-u)+t_2(1-u)^2+\cdots,  \nonumber \\
& Z=z_0-z_1(1-u)+z_2(1-u)^2+\cdots,
\end{align}
Eqs.~(\ref{eq_MP_U})-(\ref{eq_MP_Z}) lead to the condition 
\begin{align}
\label{regular_cond_EH}
u_0=t_0,   
\end{align}
and the regular solutions are uniquely determined by the three free parameters, $u_0$, $z_0$, and $z_1$. 
Since we consider linear perturbations of 
the MP AdS black hole, hereafter, we set $u_0=1$ without loss of generality. 
Therefore, there are two independent regular solutions at the event horizon: a zero-mode solution 
parametrized by $z_0$ and the $\zeta=1$ convergent solution with the eigenvector 
${\bm K}_+$ in Eq.~(\ref{non_trivial_eigen}) parametrized by $z_1$.  
  
Let us consider the behavior of the vacuum expectation value of the stress-energy tensor 
along an achronal null geodesic crossing the Cauchy horizon, 
$u_\ast=1/\kappa^2$. By introducing new coordinates, $(v, \hat{\psi})$
\begin{align} 
\label{vpsi_coordinate}
& v=t-\int\frac{\ell\sqrt{h}}{2\hat{r}_0\sqrt{u}\,\eta}\sqrt{\frac{1+\epsilon U}{1+\epsilon T}}du, \nonumber \\
& \hat{\psi}=\psi-\int\frac{\ell\sqrt{h}}{2\hat{r}_0\sqrt{u}\,\eta}(\Omega+\epsilon S)
\sqrt{\frac{1+\epsilon U}{1+\epsilon T}}du, 
\end{align}
the metric~(\ref{metric_ansatz_5MP}) takes the form 
\begin{align}
\label{Null_MP} 
& ds^2=-\frac{\hat{r}_0^2\eta}{hu}(1+\epsilon T)dv^2-\frac{\ell}{u\sqrt{uh}}\sqrt{(1+\epsilon U)(1+\epsilon T)}\,dudv
\nonumber \\
&+\frac{\hat{r}_0^2\ell^2 h}{u}(1+\epsilon Z)\left(d\hat{\psi}+\frac{\cos\theta}{2}d\phi-(\Omega+\epsilon S)dv\right)^2
\nonumber \\
&+\frac{\hat{r}_0^2\ell^2}{4u}(1+\epsilon R)(d\theta^2+\sin^2\theta d\phi^2).  
\end{align}
\vspace{-5mm} 
\begin{figure}[htbp]
  \begin{center}
\includegraphics[width=100mm]{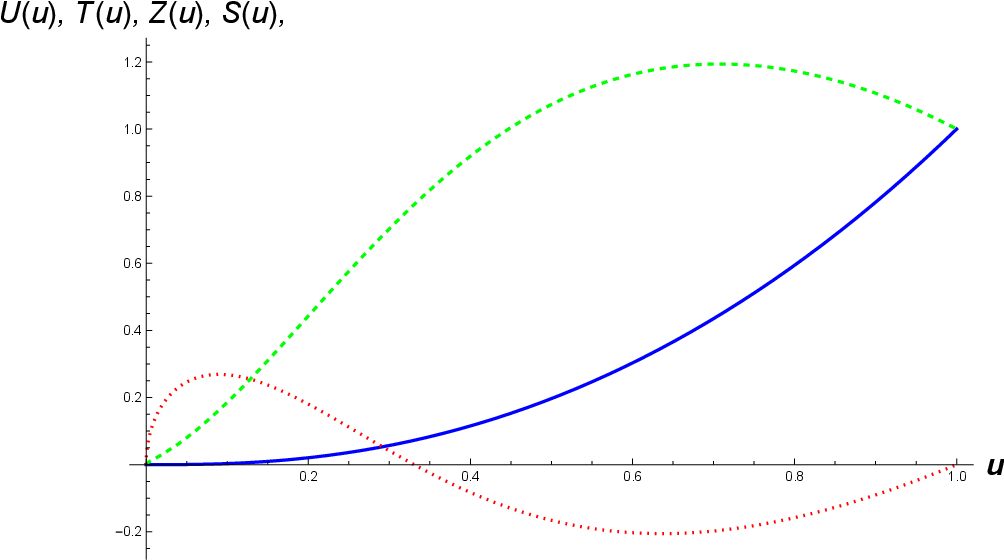}
  \end{center}
  \caption{{\small The functions $U$~(blue, solid), $T$~(green, dashes), $Z$~(red, dotted), and 
  $S$~(brown, dotdashed) are plotted in the range $u\in(0, 1)$ for the parameters, 
  $\kappa=0.5$, $\hat{r}_0=3$, $\hat{m}^2=-3$, and $(z_0,\,z_1)=(0,1)$. }}
  \label{Config_UTZS}
 \end{figure} 
 
Now, $\p_u$ is the tangent null vector along the achronal null geodesic and the affine parametrized 
tangent vector $V^\mu$ is given by 
\be
\label{null_vec_MP} 
& {\bm V}=\overline{{\bm V}}+\delta {\bm V}
=\frac{{\cal E}u^\frac{3}{2}\sqrt{h}}{\ell \sqrt{(1+\epsilon U)(1+\epsilon T)}}\frac{\p}{\p u}
\nonumber \\
&=\frac{u\sqrt{uh}\,{\cal E}}{\ell}
\left[\frac{\ell\sqrt{h}}{2\hat{r}_0\sqrt{u}\eta}\frac{\p}{\p t}+\frac{\p}{\p u}
+\frac{\ell\sqrt{h}}{2\hat{r}_0\sqrt{u}\eta}\Omega\frac{\p}{\p \psi}\right]+O(\epsilon), 
\ee
where ${\cal E}$ is a positive constant. 
The solution of~(\ref{non-extremal_NH_A}) includes the growing mode with the eigenvalue $\zeta=-1$ 
near the Cauchy horizon. The eigenvector ${\bm K}_-$ in Eq.~(\ref{non_trivial_eigen}) causes the strong 
divergence of the stress-energy tensor,    
\begin{align}
\label{null_energy_MP} 
\delta(\Exp{T_{\mu\nu}}V^\mu V^\nu)\sim \epsilon H_{\mu\nu}\overline{V}^\mu\overline{V}^\nu
=\epsilon\frac{{\cal E}^2uh(U-T)}{4\eta}=\epsilon\frac{C_1}{\lambda^2}, 
\end{align}
where $C_1$ approaches a constant near the Cauchy horizon, and $\lambda~(\sim 1/\kappa^2-u)$ is the affine 
parameter, which is defined to be zero at the Cauchy horizon. 

Fig.~\ref{Config_UTZS} shows the behavior of the functions $U$~(blue, solid), $T$~(green, dashed), $Z$~(red, dotted), 
and $S$~(brown, dotdashed) in the range $u\in(0, 1)$ for the parameter values, $\kappa=0.5$, 
$\hat{r}_0=3$, $\hat{m}^2=-3$, and $(z_0,\,z_1)=(0,1)$: The function $U$ decays as $U\sim u^{3/2}$, while the other 
functions $T$, $Z$, and $S$ decays as $\sim u^{1/2}$, as shown in Eqs.~(\ref{asymp_t_mode}) and (\ref{asymp_z_mode}). 

\begin{figure}[htbp]
 \begin{minipage}{0.47\hsize}
  \begin{center}
   \includegraphics[width=65mm]{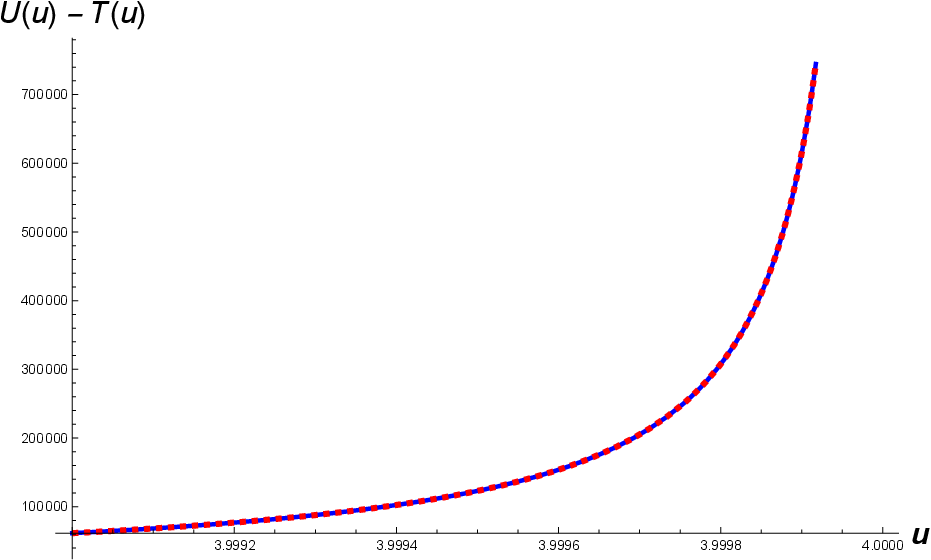}
  \end{center}
  \caption{{\small The plot of $U-T$ for the parameters $\kappa=0.5$, $\hat{r}_0=3$, $\hat{m}^2=-3$, 
  and $(z_0,\,z_1)=(0,1)$. The 
  best-fit curve~(dashed curve) is $U-T\simeq 61.5(1/\kappa^2-u)^{-1}-24.6$. }}
  \label{U_T_div_05_01_3}
 \end{minipage}
 \begin{minipage}{0.45\hsize}
  \begin{center}
   \includegraphics[width=60mm]{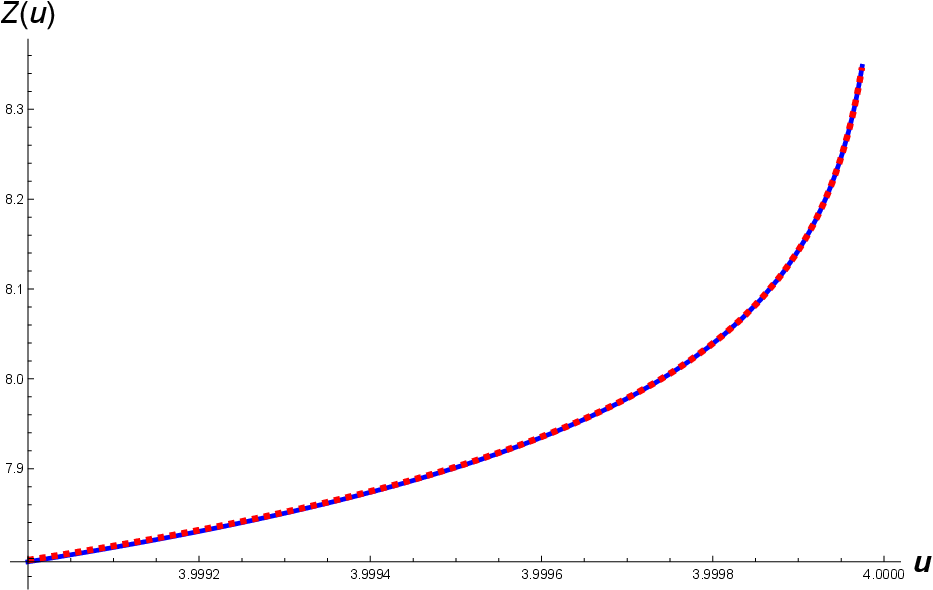}
  \end{center}
  \caption{{\small The plot of $Z$ for the parameters $\kappa=0.5$, $\hat{r}_0=3$, $\hat{m}^2=-3$, and $(z_0,\,z_1)=(0,1)$. 
   The best-fit curve~(dashed curve) corresponds to $Z\simeq -0.15\ln(1/\kappa^2-u)+6.76$. }}
  \label{Z_div_05_01_3}
 \end{minipage}
\end{figure}

Inside the event horizon, the perturbation generally grows toward the Cauchy horizon for most of the parameter 
space $(z_0, z_1)$. 
As indicated in the linear analysis~(\ref{non-extremal_NH_A}) and (\ref{non_trivial_eigen}), 
$U-T$ and $Z$ diverge near the Cauchy horizon as
\begin{align}
\label{UZ_CH}
U-T\sim \frac{C_1}{1/\kappa^2-u}, \qquad 
Z\sim C_2\ln (1/\kappa^2-u), 
\end{align}
as shown in Figs.~\ref{U_T_div_05_01_3} and \ref{Z_div_05_01_3} for the same parameters with Fig.~\ref{Config_UTZS}. 
The former divergence leads to a strong divergence of the null energy~(\ref{null_energy_MP}). As demonstrated in a 
quantum field theory on the Reissner-Nordstr\"{o}m de Sitter background~\cite{Hollands:2019whz}, one expects 
that the coefficient $C_1$ always takes a nonzero value in higher-dimensional spacetimes, except in the 
three-dimensional case analyzed in the previous section.

\begin{figure}[htbp]
 \begin{minipage}{0.47\hsize}
  \begin{center}
   \includegraphics[width=60mm]{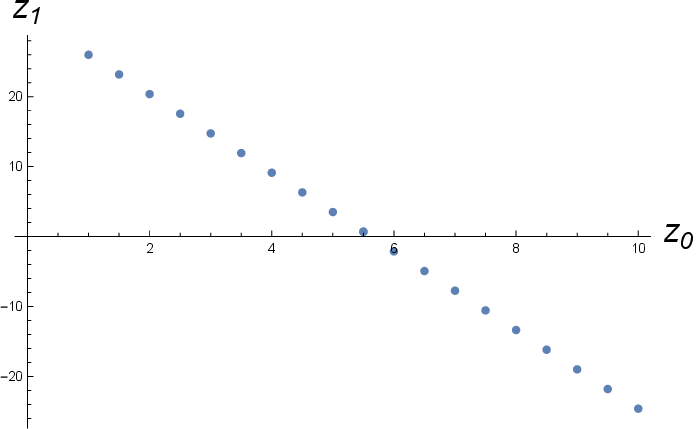}
  \end{center}
  \caption{{\small The curve $z_1=z_1(z_0)$, along which the coefficient $C_1$ vanishes, is plotted for 
  the parameters $\kappa=0.5$ and $\hat{r}_0=3$.}}
  \label{z0z1_relation}
 \end{minipage}
 \begin{minipage}{0.45\hsize}
  \begin{center}
   \includegraphics[width=60mm]{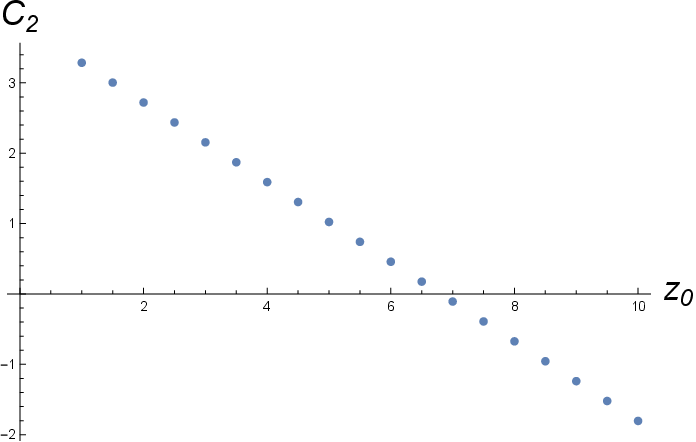}
  \end{center}
  \caption{{\small The coefficient $C_2$ is plotted along the curve $z_1=z_1(z_0)$ for the same 
  parameters $\kappa=0.5$ and $\hat{r}_0=3$.}}
  \label{C2z0_relation}
 \end{minipage}
\end{figure}

In contrast, for the five-dimensional semiclassical solution, there exists a class of parameters for which the coefficient
$C_1$ vanishes.  Fig.~\ref{z0z1_relation} shows the curve $z_1=z_1(z_0)$ in the parameter 
space $(z_0, z_1)$ where $C_1=0$, for the same background parameters as Fig.~\ref{Config_UTZS}. 
One might then consider that an observer traveling along the achronal null geodesic would not experience any divergence 
in the null energy~(\ref{null_energy_MP}), allowing the null geodesic to be extended beyond the Cauchy horizon. 
However, in our case, backreaction effects from quantum corrections are incorporated through the semiclassical equations~(\ref{semi_eqs}). 
As a result, a strong curvature singularity still forms at the Cauchy horizon, even along this curve, due to the 
presence of the logarithmic mode~(\ref{UZ_CH}) near the Cauchy horizon, as discussed below.

Let us calculate a curvature component in the frame that is parallelly propagated along the null geodesic 
with tangent vector (\ref{null_vec_MP}). A unit spacelike vector ${\bm E}_\theta$ defined by 
\be 
{\bm E}_\theta=\frac{2\sqrt{u}}{\hat{r}_0\ell\sqrt{1+\epsilon\,R}}\frac{\p}{\p \theta}
\ee
is invariant along the null geodesic orbit, i.~e.~, $V^\nu D_\nu E_\theta^\mu=0$, and orthogonal to ${\bm V}$. 
Up to O($\epsilon$), we find that the following Riemann component strongly diverges as 
\begin{align}
\delta(\bdyR_{\mu\alpha\nu\beta}{E}_\theta^\mu V^\alpha {E}_\theta^\nu V^\beta)\sim
\epsilon \frac{C_2{\cal E}^2(1+\hat{r}_0^2)(1+\kappa^2)}
{4\ell^2\kappa^8\left(\frac{1}{\kappa^2}-u\right)^2}\sim \frac{1}{\lambda^2}
\end{align}
in the presence of the logarithmic mode $Z\sim C_2\ln(1/\kappa^2-u)$. 
This is a parallelly propagated~(p.~p.)~curvature singularity~\cite{HawkingEllis} along the null geodesic 
transverse to the Cauchy horizon. The shear tensor $\sigma_{\mu\nu}$ of the null geodesic congruence, 
contracted with the unit spacelike vector ${\bm E}_\theta$, obeys the following equation along the null geodesic, 
\begin{align}
\frac{d(\sigma_{\mu\nu}E_\theta^\mu E_\theta^\nu)}{d\lambda}\sim 
\delta(C_{\mu\alpha\nu\beta}{E}_\theta^\mu V^\alpha {E}_\theta^\nu V^\beta)\sim 
\delta(\bdyR_{\mu\alpha\nu\beta}{E}_\theta^\mu V^\alpha {E}_\theta^\nu V^\beta)\sim \frac{1}{\lambda^2}
\end{align}
along the curve $z_1=z_1(z_0)$ with $C_1=0$. Here, we have used the fact that 
$\bdyR_{\mu\nu}V^\mu V^\nu\sim \Exp{T_{\mu\nu}}V^\mu V^\nu$ is subdominant compared to the 
divergence of the Riemann tensor component by Eqs.~(\ref{semi_eqs}). 
Thus, the shear tensor itself diverges as $1/\lambda$. Since the integral of the shear tensor also diverges, 
the p.~p.~singularity corresponds to a strong curvature singularity, characterized by an infinite stretching 
or crushing of the null geodesic congruence.

Fig.~\ref{C2z0_relation} illustrates that the coefficient $C_2$ is generically non-zero along the curve 
$z_1=z_1(z_0)$, except at a particular point $z_0=z_0^\ast\simeq 6.81$ for $(\kappa, \hat{r}_0)=(0.5,\,3)$. 
Contrary to the naive expectation that quantum fluctuations would universally destroy the Cauchy horizon for 
any semiclassical solutions, we find that a $C^0$-regular Cauchy horizon exists at a specific point where 
the perturbation remains finite. 
This occurs due to the fact that the solutions of~(\ref{non-extremal_NH_A}) include only two growing modes, 
which can be eliminated by a suitable adjustment of the two free parameters $(z_0, z_1)$ on the regular event horizon.

\subsection{The extremal solution}
In this subsection, we investigate semiclassical $\kappa=1$ extremal solutions of Eqs.~(\ref{eq_MP_U}), 
(\ref{eq_MP_T}), and (\ref{eq_MP_Z}). Near the extremal horizon, $u=1$, defining $P$, $Q$, $R$, and 
${\bm V}$ as $P=(1-u)U'$, $Q=(1-u)Z'$, $R=(1-u)T'$, and ${\bm V}=(U,\,Z,\,T,\,P,\,Q,\,R)$, respectively, 
${\bm V}$ obeys   
\begin{align}
\label{ext_MP_matrix}
(1-u)\frac{d{\bm V}}{du}\simeq A{\bm V}, \qquad A:=
\begin{pmatrix}
0 & 0 & 0  & 1  & 0  &  0 \\
0 & 0 & 0 & 0 & 1 & 0 \\
0 & 0 & 0 & 0 & 0 & 1 \\
\frac{\hat{r}_0^2(\hat{m}^2-16)-8}{4+12\hat{r}_0^2} & \frac{2+4\hat{r}_0^2}{1+3\hat{r}_0^2} & \frac{2\hat{r}_0^2}{1+3\hat{r}_0^2} & 3 & 0 & 0 
\\
\frac{\hat{r}_0^2(\hat{m}^2+8)+4}{2+6\hat{r}_0^2} & \frac{24+(32+\hat{m}^2)\hat{r}_0^2}{4+12\hat{r}_0^2} & 
\frac{2+4\hat{r}_0^2}{1+3\hat{r}_0^2} & 2 & 1 & 2 \\
 \frac{\hat{r}_0^2(8-\hat{m}^2)+4}{2+6\hat{r}_0^2} & -2+\frac{2\hat{r}_0^2}{1+3\hat{r}_0^2} & 
\frac{(\hat{m}^2-8)\hat{r}_0^2}{4+12\hat{r}_0^2} & -4 & 0 & -1 
\end{pmatrix}
. 
\end{align}
Substituting ${\bm V}=(1-u)^\zeta {\bm K}$ into Eq.~(\ref{ext_MP_matrix}), we arrive at the eigenvalue 
equation $|A+\zeta I|=0$, where $\zeta$ represents the eigenvalue. The solutions to this equation are determined by
\begin{align} 
\label{def_k}
& k(z)=0, \nonumber \\
& k(z):=64(1+3\hat{r}_0^2)^3z^3-16(1+3\hat{r}_0^2)^2\{35-\hat{r}_0^2(65-3\rho)\}z^2 \nonumber \\
&+4(1+3\hat{r}_0^2)\{259+\hat{r}_0^2(1154-38\rho)+3\hat{r}_0^4(377-(22-\rho)\rho)  \}z \nonumber \\
&-\{15+\hat{r}_0^2(21+\rho) \}\{15+\hat{r}_0^2(18+\hat{r}_0^2(27-\rho))(5+\rho) \}, 
\end{align}
where $z$ and $\rho$ are defined as $z:=(\zeta+1/2)^2$ and $\rho=-\hat{m}^2>0$, respectively. Here, $k(z)$ is a cubic polynomial whose coefficients depend on the parameters $\hat{r}_0$ and $\rho$.

We demonstrate that there exist at least two independent, convergent solutions with positive $\zeta$, and every solution of $k(z)=0$ satisfies $\zeta < 2$ near the extremal horizon. This ensures that the solutions remain bounded near the extremal horizon, consistent with the expected physical behavior in this regime. The result is summarized in the following proposition:\\
\\
{\it Proposition} \\
There exist at least two independent, convergent solutions with positive $\zeta$ to the coupled first-order 
differential equations (\ref{ext_MP_matrix}) near the extremal horizon. Furthermore, all solutions satisfy $\zeta< 2$ 
within the mass range specified by (\ref{mass_inequality_5D}). \\
{\it Proof}) \\
By Lemma 1 in Appendix~\ref{appendix_B}, the function $k(z)$ has a maximum value $k_{\text{max}}$ at $z = z_-$ and a minimum value $k_{\text{min}}$ at $z = z_+$, where $z_\pm$ satisfy $1/4 < z_- < z_+ < 25/4$ in the range (\ref{mass_inequality_5D}).
From Lemma 2 in Appendix~\ref{appendix_B}, we have $k_{\text{max}} > 0$ and $k_{\text{min}} < 0$ within this range. Furthermore, it is straightforward to verify that $k(25/4) > 0$ in the same range (\ref{mass_inequality_5D}).
By the intermediate value theorem, $k(z)$ has at least two distinct roots, corresponding to convergent solutions with positive $\zeta$, and all solutions satisfy $\zeta< 2$. \qquad $\Box$

For the mode solutions, the asymptotic AdS boundary condition $S(0) = 0$ is satisfied, as shown in the previous section. 
Consequently, the above proposition implies the existence of at least two extremal solutions that converge to 
the extremal Myers-Perry (MP) AdS black hole at the extremal horizon. 

Although the extremal hairy black hole solution approaches the extremal MP AdS black hole at the horizon, the stress-energy tensor $\Exp{T_{\mu\nu}}$ near the event horizon diverges along the null geodesic with the tangent vector given in Eq.(\ref{null_vec_MP}), similar to the $3$-dimensional case discussed in Sec.\ref{extremal_BTZ}. Contracting $\Exp{T_{\mu\nu}}$ with the null vector (\ref{null_vec_MP}) and using the eigenvector ${\bm K}$ that satisfies $|A +\zeta I| = 0$, we find
\begin{align} 
& \delta(\Exp{T_{\mu\nu}}V^\mu V^\nu) 
\sim \frac{\epsilon \gamma_5 w_5}{16\pi G_5 \ell^2} \frac{{\cal E}^2 u h(U - T)}{4 \eta} \sim
(1 - u)^{\zeta- 2}. 
\end{align}
Thus, all semiclassical extremal solutions must exhibit a singularity at the horizon.

\section{Summary and discussions}
\label{sec:5}
 We investigated odd-dimensional holographic semiclassical rotating AdS black holes within the framework of the AdS/CFT duality, considering a strongly coupled quantum field. Within the regime of linear perturbations, the backreaction 
of the quantum field on the geometry is incorporated via the semiclassical Eqs.~(\ref{semi_eqs}) on the conformal AdS boundary. The separation constant $m^2$ in Eqs.~(\ref{H_equation}) is constrained within the ranges given in 
Eqs.~(\ref{mass_inequality_3D}) and (\ref{mass_inequality_5D}), as shown in 
Refs.~\cite{Ishibashi:2023luz, Ishibashi:2024fnm}. Thus, the only remaining task for constructing 
the semiclassical solutions is to solve the massive Kaluza-Klein graviton equations~(\ref{H_equation}) with a negative 
mass $m^2$.  
 
In the $3$-dimensional case, we analytically derived both non-extremal and extremal semiclassical black hole 
solutions. One key feature is that the vacuum expectation value of the stress-energy 
tensor $\Exp{T_{\mu\nu}}$ always diverges near the Cauchy horizon, independent of the parameters of the 
background BTZ solution. This divergence contrasts with the regular behavior of 
$\Exp{T_{\mu\nu}}$ on the Cauchy horizon for a quantum scalar field in a fixed near-extremal BTZ black hole background~\cite{Dias:2019ery}. As pointed out in~\cite{Dias:2019ery}, this finite behavior arises due to a 
special property of the background BTZ geometry. On the other hand, in our semiclassical model, the vacuum expectation value of the stress-energy tensor is computed on the modified BTZ geometry via the semiclassical equations. 
Since this special property is absent in the modified geometry, we infer that the vacuum expectation value 
generically diverges on the Cauchy horizon in the semiclassical theory.        
 Another key feature in the non-extremal case is that the degree of divergence depends on the type of perturbation, corresponding to even- or odd-parity modes in the static frame of the background BTZ solution. For even-parity perturbations, the null energy along a null geodesic with tangent vector $V^\mu$ exhibits a strong divergence as $\Exp{T_{\mu\nu}}V^\mu V^\nu \sim 1/\lambda^2$ with respect to the affine parameter $\lambda$ ($\lambda = 0$ at the Cauchy horizon). Conversely, for odd-parity perturbations, the divergence is milder, scaling as 
$\Exp{T_{\mu\nu}}V^\mu V^\nu \sim 1/\lambda$. These findings indicate that a $C^0$ extension of the metric beyond the Cauchy horizon is feasible in the absence of even-parity perturbations, although generic perturbation supports the SCC 
conjecture. 

In the 5-dimensional case, semiclassical non-extremal rotating AdS black holes are characterized by two parameters at the regular event horizon. Although $\Exp{T_{\mu\nu}}V^\mu V^\nu$ typically diverges strongly as $\sim C/\lambda^2$ 
near the Cauchy horizon for most parameter choices, there exists a class of semiclassical solutions in which the coefficient $C$ vanishes, independent of the background MP black hole parameters. Even in such cases, we demonstrated 
that a parallely propagated Riemann tensor component along the null geodesic still generically diverges strongly as
 $\sim C/\lambda^2$. This appears to be a peculiar phenomenon observed only in the semiclassical model, which 
incorporates the backreaction on the geometry and it supports the SCC conjecture. 

In contrast, there exist semiclassical solutions 
with a $C^0$-regular Cauchy horizon where perturbations remain finite.
This is somewhat surprising, as one might expect quantum effects to universally induce strong divergences through backreaction, independent of the background geometry. Unlike in the 3-dimensional case, only two of the six independent mode solutions contribute to the singular behavior at the Cauchy horizon. By carefully tuning the two free parameters at the event horizon, a semiclassical solution with a $C^0$-regular Cauchy horizon can be constructed. In scenarios 
with a fixed curved background, such strong divergence has been observed in free quantum fields on fixed 4-dimensional 
Reissner-Nordstr\"{o}m-dS black hole backgrounds~\cite{Hollands:2019whz} and in strongly coupled field theories on fixed 5-dimensional Myers-Perry AdS black hole spacetimes~\cite{Ishibashi:2017jde}. In both cases, the null energy density exhibits a strong divergence near the Cauchy horizon in the absence of backreaction effects. However, when backreaction is incorporated, it has the potential to weaken the divergence, suggesting that quantum effects could moderate the singularity under semiclassical perturbations.

For the extremal case, we examine semiclassical solutions under linear perturbations. Notably, we obtained analytic 3-dimensional semiclassical black hole solutions that converge to the extremal BTZ solution at the event horizon. Since 
the decay rate is very slow, the null energy $\Exp{T_{\mu\nu}}V^\mu V^\nu$ diverges 
along a null geodesic with tangent vector $V^\mu$ transverse to the event horizon. In the 5-dimensional hairy AdS 
black hole case, there exist at least two mode solutions that vanish at the extremal horizon. 
We proved that the decay rate for the mode solutions is very slow at the extremal horizon, and consequently 
$\Exp{T_{\mu\nu}}V^\mu V^\nu$ always diverges along a null geodesic crossing the event horizon for any 
extremal semiclassical solution. 
These results indicate that such singularities at the event horizon are a generic feature of semiclassical rotating extremal solutions, similar to the findings for classical perturbations~\cite{Maeda:2011pk, Iizuka:2022igv, Horowitz:2022mly, Horowitz:2024kcx}. Whether this generic singular behavior extends to semiclassical charged extremal black hole solutions, as 
suggested in the s-wave approximation~\cite{Arrechea:2021ldl}, remains an open question.

Finally, the thermodynamic stability of these semiclassical rotating black hole solutions warrants further investigation. In the non-rotating case, 3- and 5-dimensional semiclassical hairy black hole solutions are thermodynamically more stable than the BTZ or hyperbolic AdS black hole solutions with a vanishing stress-energy tensor $\Exp{T_{\mu\nu}} = 0$~\cite{Ishibashi:2023psu, Ishibashi:2024fnm}. Furthermore, extending semiclassical solutions beyond the linear perturbation regime will be crucial for a deeper understanding of the backreaction effects. A detailed analysis of the rotating case will be pursued in future work.

\begin{center}
{\bf Acknowledgments}
\end{center}
We wish to thank Takashi Okamura and Akihiro Ishibashi for useful discussions.  
This work is supported in part by JSPS KAKENHI Grant No. 20K03975 (K.M.) and also supported by MEXT KAKENHI 
Grant-in-Aid for Transformative Research Areas A Extreme Universe No.21H05186 (K.M.). 

\appendix
\section{$d=5$ Semiclassical equations}
\label{appendix_A}
Under the transverse and traceless condition~(\ref{TT_cond}), the first order perturbation of the 
following curvature tensors and the derivatives are given by 
\begin{align}
\label{d=5_pert_curvature}
& \delta\bdyR_{\mu\nu}=-\epsilon\left(\frac{1}{2}\overline{\bdyD}^2H_{\mu\nu}
+\frac{4}{\ell^2}H_{\mu\nu}+\overline{\bdyR}_{\mu\alpha\nu\beta}H^{\alpha\beta}\right)
=-\epsilon\left(\frac{m^2}{2}+\frac{4}{\ell^2}  \right)H_{\mu\nu}, \nonumber \\
& \delta(\bdyD^2\bdyR_{\mu\nu})=-\frac{\epsilon}{2}m^2\overline{D^2}H_{\mu\nu}, \qquad 
\delta(\bdyR \bdyR_{\mu\nu})=\frac{20\epsilon}{\ell^2}\left(\frac{4}{\ell^2}+\frac{m^2}{2}\right)H_{\mu\nu}, 
\nonumber \\
& \delta(\bdyR_{\alpha\mu\beta\nu}\bdyR^{\alpha\beta})=
\frac{4\epsilon}{\ell^2}\left(\frac{m^2}{2}+\frac{4}{\ell^2}  \right)H_{\mu\nu}
-\frac{\epsilon}{2}m^2\overline{\bdyR}_{\mu\alpha\nu\beta}H^{\alpha\beta}, \nonumber \\
& \delta\bdyR=\delta(\bdyD_\mu \bdyD_\nu \bdyR)=\delta(\bdyR_{\mu\nu}\bdyR^{\mu\nu})=0,  
\end{align} 
where we used Eq.~(\ref{H_equation}) in the second equality in the first line. Substituting this 
into the perturbation of $\tau_{\mu\nu}^{(5)}$ in (\ref{vev_SE_ct}), one obtains the same 
tensor~\cite{Ishibashi:2024fnm}
\begin{align}
\label{tau_d=5}
\delta \tau^{(5)}_{\mu\nu}=-\frac{\epsilon L^4}{72\pi\ell G_6\sin\frac{z}{\ell}}  
\left(\frac{m^4}{2}+\frac{9m^2}{4\ell^2}-\frac{9}{2\ell^4} \right)H_{\mu\nu}. 
\end{align}
Inserting Eq.~(\ref{tau_d=5}) into (\ref{vev_SE_tensor}) and expanding (\ref{xi_sol}) near $z=0$, 
one obtains the perturbed stress-energy tensor~\cite{Ishibashi:2024fnm} 
\begin{align}
& \delta \Exp{T_{\mu\nu}}=-\frac{\epsilon}{16\pi G_5}\frac{\gamma_5 w_5(\hat{m}^2)}{\ell^2}H_{\mu\nu}, \nonumber \\
&  w_5(\hat{m}^2)=\frac{\pi}{9}\frac{\hat{m}^2(3+\hat{m}^2)\sqrt{4+\hat{m}^2}}{\tan(\pi\sqrt{4+\hat{m}^2})}. 
\end{align}
By Eqs.~(\ref{d=5_pert_curvature}), the perturbation of Eqs.~(\ref{semi_eqs}) becomes 
\begin{align}
\label{d=5_semi_eqs}
-\epsilon\frac{m^2}{2}H_{\mu\nu}=-\frac{\epsilon}{16\pi G_5}\frac{\gamma_5 w_5(\hat{m}^2)}{\ell^2}H_{\mu\nu}, 
\end{align}
which is equivalent to the algebraic equation~\cite{Ishibashi:2024fnm}
\begin{align}
\label{algebra_5D}
\gamma_5=\frac{\hat{m}^2}{w_5(\hat{m}^2)}. 
\end{align}

\section{Lemmas}
\label{appendix_B}
\noindent {\it Lemma 1} \\
The function $k(z)$ in (\ref{def_k}) takes the maximum and the minimum values at $z=z_-$ and $z=z_+$, 
respectively. $z_\pm$ satisfies the inequality, 
\begin{align}
\frac{1}{4}<z_-<z_+<\frac{25}{4}. 
\end{align}
{\it proof}) \\
By $k'(z)=0$, $z_\pm$ is given by 
\begin{align} 
z_\pm=\frac{1}{36}\left[65-3\rho+\frac{40+3\rho}{1+3\hat{r}_0^2}\pm 
\frac{12\sqrt{28+\hat{r}_0^2(68-6\rho)+\hat{r}_0^4(52-12\rho)}}{1+3\hat{r}_0^2}\right]
=:\frac{p_1\pm 12\sqrt{q_1}}{36(1+3\hat{r}_0^2)}, 
\end{align}
where $q_1>0$ in the range~(\ref{mass_inequality_5D}) for an arbitrary value of $\hat{r}_0$. 
By
\begin{align}
z_--\frac{1}{4}=\frac{32+\hat{r}_0^2(56-3\rho)-4\sqrt{q_1}}{12(1+3\hat{r}_0^2)}=:
\frac{p_2-4\sqrt{q_1}}{12(1+3\hat{r}_0^2)}, 
\end{align}
and $p_2>0$ in the range~(\ref{mass_inequality_5D}),  
the inequality
\begin{align} 
\label{inequality_1}
p_2^2-16q_1=576+96\hat{r}_0^2(26-\rho)+9\hat{r}_0^4\left\{(\rho-8)^2+192\right\}>0
\end{align}
shows $z_1>1/4$. Similarly, one finds 
\begin{align}
\frac{25}{4}-z_+=\frac{40+\hat{r}_0^2(160+3\rho)-4\sqrt{q_1}}{12(1+3\hat{r}_0^2)}=\frac{p_3-4\sqrt{q_1}}{12(1+3\hat{r}_0^2)}, 
\end{align}
where $p_3>0$. The inequality (\ref{inequality_1}) combined with  
\begin{align} 
p_3^2-16q_1=1152+48\hat{r}_0^2(244+7\rho)+9\hat{r}_0^4\{2752+\rho(128+\rho)\}>0
\end{align}
proves the above lemma 1. \\
\\
{\it Lemma 2} \\
The maximum and minimum values of $k(z)$ are positive and negative, respectively.  \\
{\it proof}) \\
The maximum and minimum values of $k(z)$ are given by 
\begin{align} 
\label{extreme_k}
k(z_\pm)=\frac{256}{27}\left[2(1+\hat{r}_0^2)\{2\hat{r}_0^4(35-9\rho)+\hat{r}_0^2(5-9\rho)-20  \}
\mp\{\hat{r}_0^4(26-6\rho)+\hat{r}_0^2(34-3\rho)+14 \}\sqrt{q_1}\right]. 
\end{align}
The minimum value $k(z_+)$ is maximized at 
\begin{align}  
\rho_0=3+\frac{2}{\hat{r}_0^2}-\frac{1}{1+2\hat{r}_0^2}>3
\end{align}
with respect to $\rho$.
When $\rho_0\le 15/4$, 
\begin{align} 
k(z_+)\Bigr{|_{\rho=\rho_0}}=-512(1+\hat{r}_0^2)(2+3\hat{r}_0^2)<0.  
\end{align}
When $\rho_0>15/4$, 
\begin{align}
k(z_+)\Bigr{|_{\rho=\frac{15}{4}}}&=-\frac{32}{27}\left[20(16+39\hat{r}_0^2+21\hat{r}_0^4-2\hat{r}_0^6)
+7\sqrt{14}(8+13\hat{r}_0^2+2\hat{r}_0^4)^\frac{3}{2} \right] \nonumber \\
\qquad &<-\frac{32}{27}\left[20(16+39\hat{r}_0^2+21\hat{r}_0^4)+(14\sqrt{28}-40)\hat{r}_0^6 \right]<0, 
\end{align}
showing $k(z_+)<0$ in the range~(\ref{mass_inequality_5D}). Here, we used 
$7\sqrt{14}(8+13\hat{r}_0^2+2\hat{r}_0^4)^\frac{3}{2}>14\sqrt{28}\,\hat{r}_0^6$ in the first inequality in the second line. 
 
On the other hand, the maximum value $k(z_-)$ is a monotonously decreasing function of $\rho$, and hence, 
\begin{align} 
& k(z_-)\Bigr{|_{\rho=\frac{15}{4}}}=-\frac{640}{27}(1+\hat{r}_0^2)(16+23\hat{r}_0^2-2\hat{r}_0^4)
+\frac{224}{27}\sqrt{14}(8+13\hat{r}_0^2+2\hat{r}_0^4)\sqrt{8+13\hat{r}_0^2+2\hat{r}_0^4} \nonumber \\
&>-\frac{640}{27}(1+\hat{r}_0^2)(16+23\hat{r}_0^2-2\hat{r}_0^4)
+\frac{224}{27}\sqrt{28}(8+13\hat{r}_0^2+2\hat{r}_0^4)(2+\hat{r}_0^2) \nonumber \\
&=\frac{64}{27}\left[
16(7\sqrt{7}-10)+(238\sqrt{7}-390)\hat{r}_0^2+(119\sqrt{7}-210)\hat{r}_0^4+(20+14\sqrt{7})\hat{r}_0^6\right]>0,  \qquad 
\Box
\end{align} 
where we used $8+13\hat{r}_0^2+2\hat{r}_0^4>8+8\hat{r}_0^2+2\hat{r}_0^4$ in the second inequality.


\end{document}